\documentclass[aps,showpacs,12pt]{revtex4} 
\usepackage{psfig}

\begin{document}
\title{Comment on cond-mat/0107371:  ``Dynamical exponents of an 
  even-parity-conserving contact process with diffusion''}

\author{M. Argollo de Menezes$^\dagger$ and Ronald Dickman$^\star$}

\affiliation{$^\dagger$ Department of Physics, University of Notre Dame,
  Notre Dame, In 46556\\ $^\star$ Departamento de F\i{\i}sica, ICEx,
  Universidade Federal de Minas Gerais, Caixa Postal 702,\\ 30161-970
  Belo Horizonte, MG, Brazil }

\date{\today}
%
%

\begin{abstract}
  In cond-mat/0107371, de Mendon\c{c}a proposes that diffusion can
  change the universality class of a parity-conserving reaction-diffusion
  process. In this comment we suggest that this cannot happen, due to symmetry
  arguments. we also present numerical results from lattice simulations which
  support these arguments, and mention a previous result supporting
  this conclusion.
\end{abstract}

\pacs{05.40.+j, 05.70.Ln } 

\maketitle


Although there are a large variety of nonequilibrium
reaction/reaction-diffusion models which present a transition from an
active to an absorbing state, their critical behaviors fall within a
small number of universality classes. The simplest one, which can be
viewed as the nonequilibrium counterpart of the Ising universality
class, is the contact process (CP), or directed percolation (DP)
universality class ~\cite{harris,grassberger,liggett}. Another
well-established class is the parity-conserving (PC) universality
class ~\cite{tretyakov,pc-cardy}. More recently, binary spreading (BS)
processes (with or without parity conservation), have appeared as
representative of a new universality class ~\cite{tauber,pcpd,binary}.

It appears that the nature of the creation and annihilation processes is
fundamental on determining to which universality class a given process might
belong, that is, the number of particles necessary for a reaction
to occur seems to play a key role. One way to assess the importance of such
terms is via numerical simulations or, whenever possible, by
studying the analytical properties of the equations governing the dynamical
process.

For any Markovian process, one can write a master equation describing
how the probability $P(\{a\},t)$ of a given configuration $\{a\}$
evolves in time ~\cite{dickman}.  A technique first introduced by Doi
~\cite{doi} allows us to derive a field-theoretic representation for
the master equation, and then to handle fluctuations
systematically. Moreover, with the field-theoretic action in hand,
one can write down the Langevin equation which governs the stochastic
evolution of the field operators, with no ambiguity on the definition
of the noise~\cite{tauber}.

Even though the noise is always multiplicative in absorbing state phase
transitions, it can have different functional forms, which can change entirely
the long-term behavior of the system being described.  For instance, when two
particles are necessary for annihilation to occur, the noise in the Langevin
equation is, counter-intuitively, purely \emph{imaginary} (as in
representatives of the PC class). On the other hand, when both creation and
annihilation require a pair of particles, there is a competition between
``real'' and ``imaginary'' noise (as in representatives of the BS
class)\protect\footnote{Systems where a pair of particles are needed for
  creation of a third one, and single particle annihilation, belong to the
  same universality class as DP~\cite{rigid}}.

\vspace{1cm}
\begin{table}[!h]
  \begin{center}
    \begin{tabular}{|c|c|c|} 
      \hline
      Universality class & creation  & annihilation \\
      \hline
      DP   & unary  & unary \\
      PC   & unary  & binary \\
      PCPD & binary & binary \\
      \hline
    \end{tabular}
    \caption{Nature of creation/annihilation processes within each
      universality class, that is, the number of particles needed for
      a reaction to occur.}
  \end{center}
\end{table}

The CP(M) process, invented by Inui and Tretyakov ~\cite{cpm}, is a
parity-conserving dynamical process that can be summarized as
follows

\begin{eqnarray}
\mbox{mA}&\to& \emptyset \nonumber \\
\mbox{A}&\to& \mbox{(m+1)A}.
\end{eqnarray}

It can be interpreted as a process where a particle generates $m$
others with probability $p$, while $m$ particles, upon contact, are
annihilated with probability $1-p$. An isolated cluster of $(m-1)$
particles cannot diffuse or annihilate. It has been shown by the
authors that CP(2), starting with an even number of particles, belongs
to the PC universality class ~\cite{cpm}.

Recently, Mendon\c{c}a~\cite{zericardo} studied the CP(2) process with
 diffusion of solitary particles. On the basis of finite size scaling
 of the exact numerical diagonalization of the evolution operator, he
 reports that diffusion changes the universality class from PC to DP.
 Numerical estimates for $z$, the dynamic exponent governing spatial
 fluctuations of the cluster of alive sites, seem to converge to
 $z_{DP}$.

At a first glance, the claim that the addition of diffusion might
alter the dynamical critical behavior of the CP(2) process does not
seem plausible, since it does not change the symmetry of the problem:
the number of particles is still conserved modulo $2$ and the
absorbing state is still unique and devoid of particles. Moreover, as
in the analysis of simpler models ~\cite{dickman-diffusion}, it can be
seen that a coarse-grained description of non-diffusive lattice models
generally includes a diffusive term, and even without coarse-graining,
an effective diffusion is already present in CP(2), for instance, in
the sequence $A\emptyset\emptyset \to AAA \to \emptyset\emptyset A$.

To verify this, we have developed a probabilistic cellular automaton
for the $1d$ CP(2)d (with diffusion rate $d=0.05$) with even and odd sites
constituting sub-lattices $A$ and $B$ ~\cite{details}.  Starting from a
fully occupied lattice of $8000$ sites, we have measured the density
of active sites $\rho(t)$, which should decay as $\rho(t) \propto
t^{-\delta}$ at criticality. From the data we find $\delta \sim
0.286(5)$.

We have also performed dynamical simulations at the critical point,
starting from a single pair of particles and measuring the accumulated
number of active sites, $M(t) \propto t^{\eta+1}$, from which we find
$\eta \sim 0.0(5)$. These results suggest that diffusion does no alter
the critical behavior of CP(2), and only confirm what has already been
obtained by Zhong and ben-Avraham, who studied a
branching-annihilating random walk with finite annihilation rate
~\cite{zhong}. This system is equivalent to the CP(2)d with high
diffusion probability, and thus should have the same critical
behavior. We point out to cond-mat/0207720 for recent simulations of
the diffusive pair contact process~\cite{tauber,pcpd}, a lattice model which
(possibly) has a diffusion dependent universality class.

\section*{Acknowledgments}
This work has initiated during a visit to UFMG. I thank R. Dickman
introducing me to the field of nonequilibrium phase transitions. This
work was partially supported by CNPq and FAPERJ.

\vskip 1.5cm
\noindent $^\dagger$ marcio@if.uff\\
$^\star$ dickman@fisica.ufmg.br

\newpage

\noindent FIGURE CAPTIONS
\vspace{1em}

\noindent FIG. 1. Density of active sites $\rho(t)$ as a function of time (in Monte-Carlo
  steps). Results are shown for the diffusionless CP(2) (upper) and
  CP(2)d with d=0.05 (lower).

\noindent FIG. 2. Cummulative number of active sites, $M(t)$ as a function of
  time (in Monte Carlo steps) for the CP(2) (upper) and CP(2)d
  (lower) with d=0.05.

\newpage

\noindent FIGURES

\begin{figure}[!h]
  \vspace{1cm}
{\centerline{\psfig{figure=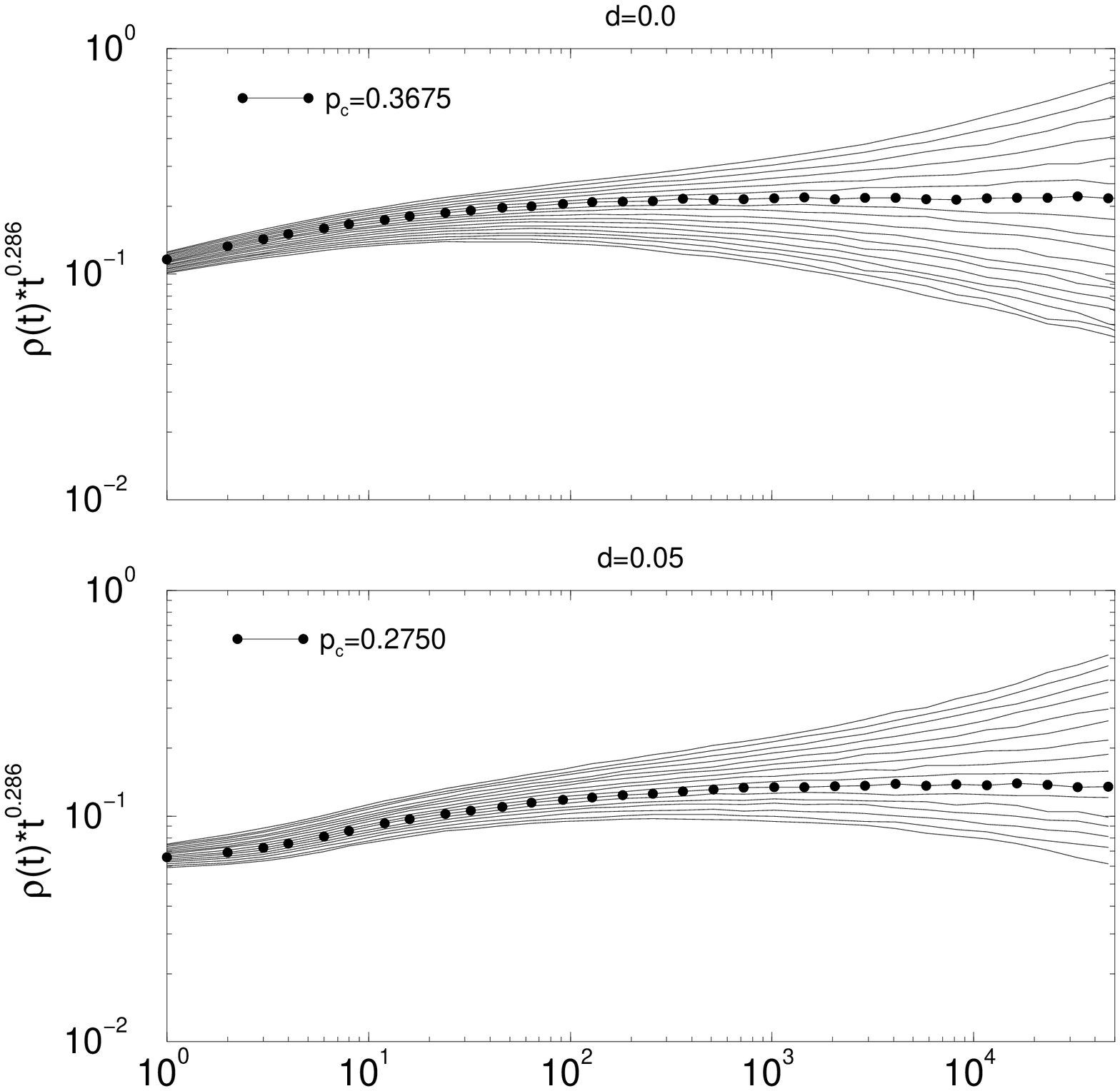,angle=270,width=8.5cm}}}
\caption{}
\end{figure}

\begin{figure}[!h]
  \vspace{1cm}
{\centerline{\psfig{figure=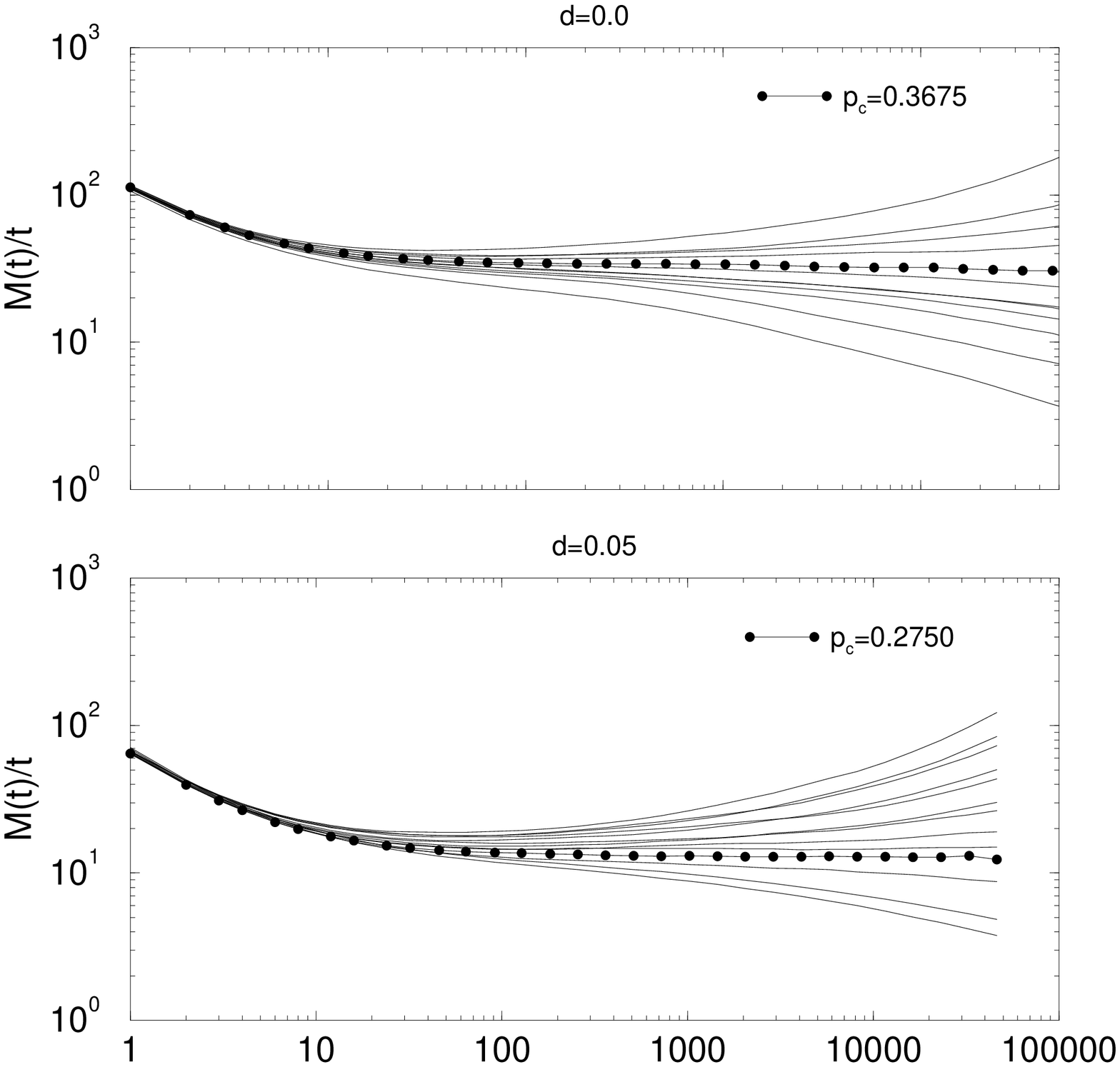,angle=270,width=8.5cm}}}
\caption{}
\end{figure}


\begin{thebibliography}{99}
\bibitem{details} A detailed explanation of the multi-spin coded algorithm and
  more numerical results will be presented elsewhere.
\end{thebibliography}
\end{document}